\documentclass[english,a4paper,manuscript]{revtex4}
\usepackage[T1]{fontenc}
\usepackage[latin1]{inputenc}
\usepackage{geometry}
\usepackage{babel}

\makeatletter

\makeatother

\begin{document}

\title{Jeans instability analysis in the presence of heat in Eckart's frame.}
\author{J. H. Mondragón Suárez, A. Sandoval-Villalbazo}
\affiliation{Departamento de Física y Matemáticas, Universidad Iberoamericana, Prolongación Paseo de la Reforma 880, México, D.F. 01219, México.}

\begin{abstract}
It is shown that the coupling of heat with acceleration first proposed by Eckart would have an overwhelming effect in the growth of density mass fluctuations, even in non-relativistic fluids in the presence of a gravitational field. Gravitational effects would be negligible if the heat-acceleration relation is assumed to be valid for the hydrodynamic equations. A direct implication of this result is that recent alternative first order in the gradients theories must be taken into account while describing a special relativistic fluid.
\end{abstract}

\maketitle

\section{Introduction}
Astronomical observations strongly suggest that structures are continuously being formed in our  Universe. According to our current knowledge, gravity is the leading interaction responsible for structure formation. The critical parameters in these events are the particle density and certain lengths that have been known since the pioneering works of J. Jeans \cite{Jeans1} \cite{Kolb}. Also, the characteristic formation time that can be inferred from these theories lead to very large values compared with those that appear in microscopic processes such as particle collisions.

In the original works regarding structure formation, the growth of linear density perturbations has been analytically
modeled in terms of the basic conservation laws of particle number and momentum. In this approach, density number perturbations satisfy a wave equation whose possible solutions provide basic values for the critical parameters that allow density growth, or simply acoustic wave propagation \cite{Kolb}. On the other hand, dissipative processes are always present in real phenomena. The effect of viscosity and heat flux on the determination of the critical parameters present in gravitational collapse have been addressed by several authors \cite{Weinberg}  \cite{Dehnen} \cite{Nos1}; most of those works use a classical (non relativistic) framework in order to include dissipation.

In the present work we show that the inclusion of the heat-acceleration coupling in the framework of relativistic hydrodynamics would yield a pathological behavior regarding the observed characteristic times in which gravitational collapse takes place. Indeed, the generic instability first found by Hiscock and Linblom \cite{HL1} and later thoroughly examined in a physical context \cite{Nos2} would have a dominating (unobserved) effect over gravity.  The paper is divided as follows, section two includes the basic equations in which Eckart's framework is developed, including weak gravitational effects;  in section three the evolution of the first order fluctuations that arises from the linearized equations is analyzed. The corresponding dispersion relation is examined in the presence of gravity in order to establish the dominant effects. Finally, section four is devoted to a discussion of the results here obtained and its corresponding implications.

\section{Relativistic transport equations in Eckart's frame}
The starting point in Eckart's formalism for a simple fluid is the particle conservation equation
\begin{equation}
N_{;\mu }^{\mu }=0 .  \label{eq:1}
\end{equation}
In Eq.(\ref{eq:1}), for a number density $n$ and hydrodynamic velocity $u^{\mu }$, $N^{\mu }=n u^{\mu }$ denotes the particle flux.
In terms of total derivatives, this equation reads
\begin{equation}
\dot{n}+n\theta =0  \label{eq:2}
\end{equation}
where $\theta =u_{;\mu }^{\mu }$.
The energy-momentum tensor balance has the form
\begin{equation}
T_{\nu ;\mu }^{\mu }=0  \label{eq:3}
\end{equation}
where
\begin{equation}
T_{\nu }^{\mu }=\frac{n\varepsilon }{c^{2}}u^{\mu }u_{\nu }+ph_{\nu }^{\mu
}+\Pi _{\nu }^{\mu }+\frac{1}{c^{2}}q^{\mu }u_{\nu }+\frac{1}{c^{2}}u^{\mu
}q_{\nu }  \label{eq:4}
\end{equation}
Here, $\varepsilon $ is the internal energy per particle, $p$ is the local pressure, and  $h_{\nu }^{\mu }=\delta _{\nu }^{\mu }+\frac{u^{\mu }u_{\nu }}{c^{2}}$ is the spatial projector with signature  $(+++-)$. The Navier tensor $\Pi _{\nu }^{\mu }$ and the heat flux $q^{\mu} $ satisfy the orthogonality relations:
\begin{equation}
u_{\mu }\Pi _{\nu }^{\mu }=u^{\nu }\Pi _{\nu }^{\mu }=0,\qquad q_{\mu
}u^{\mu }=q^{\mu }u_{\mu }=0  \label{eq:5a}
\end{equation}
Now, the projection $u^{\nu }T_{\nu ;\mu }^{\mu }=0$ leads to the internal energy balance equation,
\begin{equation}
n\dot{\varepsilon}+p\theta +u_{,\mu }^{\nu }\Pi _{\nu }^{\mu }+q_{;\mu
}^{\mu }+\frac{1}{c^{2}}\dot{u}^{\nu }q_{\nu }=0  \label{eq:8b}
\end{equation}
and, together with the expression $T_{\nu ;\mu }^{\mu }=0 $, to the equation of motion:
\begin{eqnarray}
\left( \frac{n\varepsilon }{c^{2}}+\frac{p}{c^{2}}\right) \dot{u}_{\nu
}+\left( \frac{n\dot{\varepsilon}}{c^{2}}+\frac{p}{c^{2}}\theta \right)
u_{\nu }+p_{,\mu }h_{\nu }^{\mu }+\Pi _{\nu ;\mu }^{\mu }  \nonumber \\
+\frac{1}{c^{2}}\left( q_{;\mu }^{\mu }u_{\nu }+q^{\mu }u_{\nu ;\mu }+\theta
q_{\nu }+u^{\mu }q_{\nu ;\mu }\right)  &=&0  \label{eq:6}
\end{eqnarray}
Eq. (\ref{eq:8b}) can be rewritten in terms of $n$ and $T$  using the functional hypothesis $\dot{\varepsilon}=
\left( \frac{\partial \varepsilon }{\partial n}\right) _{T}
\dot{n}+\left( \frac{\partial \varepsilon }{\partial T}\right) _{n}\dot{T}$ leading to the expression
\begin{equation}
nC_{n}\dot{T}+p \theta +u_{;\mu
}^{\nu }\Pi _{\nu }^{\mu }+q_{;\mu }^{\mu }+\frac{1}{c^{2}}\dot{u}^{\nu
}q_{\nu }=0  \label{eq:10}
\end{equation}
The entropy balance is also established by means of the functional hypothesis $
\dot{s}=\left( \frac{\partial s}{\partial n}\right) _{\varepsilon }\dot{n}
+\left( \frac{\partial s}{\partial \varepsilon }\right) _{n}\dot{\varepsilon}$, leading to the expression for the entropy production $\sigma$ \cite{Eckart}:
\begin{equation}
\sigma =-\frac{q^{\nu }}{T}\left( \frac{T_{,\nu }}{T}+\frac{T}{c^{2}}\dot{u}
_{\nu }\right) -\frac{u_{,\mu }^{\nu }}{T}\Pi _{\nu }^{\mu }  \label{eq:14}
\end{equation}
This last equation has motivated the introduction of the heat-acceleration coupling
\begin{equation}
q^{\nu }=-\kappa h_{\mu }^{\nu }\left( T^{,\mu }+\frac{T}{c^{2}}\dot{u}^{\mu
}\right)   \label{eq:17}
\end{equation}
where $\kappa$ is the thermal conductivity coefficient.
In the next section we shall analyze critically the implications of Eq. (\ref{eq:17}) using the linearized version of the set of equations (\ref{eq:2},\ref{eq:6}) and Eq. (\ref{eq:10}) in the context of special relativity, and  introducing a gravitational field.

\section{Linearized equations: exponentially growing modes}
In order to examine the behavior of small fluctuations $\delta X$ around the equilibrium values of a thermodynamical variable $X$, we shall write $X=X_{o}+\delta X$, where the subscript $o$ denotes the average value. Also, in the right hand side of Eq. (\ref{eq:6}) we will explicitly write a gravitational force, taking into account a Newtonian approximation for the covariant derivatives included in this expression. Neglecting second order terms in the fluctuations, the resulting equations in the comoving frame $\vec{u}_{0}=0$ read:
\begin{equation}
\frac{\partial}{\partial t}(\delta n)+n_{0}\delta\theta=0  \label{eq:18}
\end{equation}
\begin{eqnarray}
\frac{1}{c^{2}}\left(n_{o}\varepsilon_{o}+p_{o}\right)\frac{\partial}{\partial t}(\delta \theta)+
k T_{o} \nabla^{2}(\delta n)+n_{o} k \nabla^{2}(\delta T)-A \nabla^{2}(\delta \theta)
\nonumber \\
-\frac{\kappa}{c^{2}}\nabla^{2} \frac{\partial}{\partial t}(\delta T)-
\frac{\kappa T_{o}}{c^{4}} \frac{\partial^{2}}{\partial t^{2}}(\delta \theta)=\frac{1}{c^{2}}\left(n_{o}\varepsilon_{o}+p_{o}\right) \nabla^{2} (\delta \varphi)\label{eq:19}
\end{eqnarray}
\begin{equation}
nC_{n}\delta\dot{T}+\left(\frac{T_{o}\beta}{\kappa_{T}}\right)\delta\theta-
\kappa\left(\delta T^{,k}+\frac{T_{o}}{c^{2}}\delta\dot{u}^{k}\right)_{;k}=0.
\label{eq:20}
\end{equation}
\medskip

Use has been made of the fact that $p=p\left(n,\, T\right)$. In Eq. (\ref{eq:19}) we have introduced the gravitational potential $\varphi$, which satisfies \emph{in the weak field limit} the Poisson equation $\nabla^{2}  \delta\varphi= -4 \pi G m \delta n$. Also, the linearized expression of Eq. (\ref{eq:6}) has been written in terms of the longitudinal mode $\delta \theta$.

In the Fourier-Laplace space, the system of Eqs. (\ref{eq:18}-\ref{eq:20}) reads:
\begin{equation}
s \delta \hat{\tilde{n}} + n_{o} \delta \hat{\tilde{\theta}}=\delta \tilde{n}(\vec{q},0) \label{eq:21}
\end{equation}
\begin{eqnarray}
-q^{2} k T_{o} \delta \hat{\tilde{n}}+4 \pi G m^{2} n_{o} \delta \hat{\tilde{n}}+\frac{1}{c^{2}}\left(n_{o}\varepsilon_{o}+p_{o}\right) s \delta \hat{\tilde{\theta}}+A q^{2}\delta \hat{\tilde{\theta}}-\frac{\kappa T_{o}}{c^{4}} s^{2} \delta \hat{\tilde{\theta}}
\nonumber \\- q^{2} n_{o} k \delta \hat{\tilde{T}}+\frac{\kappa q^{2}}{c^{2}} s \delta \hat{\tilde{T}}=-\frac{\kappa T_{o}}{c^{4}} s \delta \tilde{\theta}(\vec{q},0)+\frac{1}{c^{2}}\left(n_{0}\varepsilon_{o}+p_{o}\right) \delta \tilde{\theta}(\vec{q},0)-\frac{\kappa}{c^{2}} \delta \tilde{T}(\vec{q},0)    \label{eq:22}
\end{eqnarray}
\begin{eqnarray}
k n_{o} T_{o} \hat{\tilde{\delta \theta}} -\frac{k T_{o}}{c^{2}}s \hat{\tilde{\delta \theta}}+c_{n} n_{o} s \hat{\tilde{\delta T}}+\kappa q^{2}\hat{\tilde{\delta T}}=\delta \tilde{T}(\vec{q},0)+\frac{D_{th} T_{o}}{c^{2}}\delta \tilde{\theta}(\vec{q},0)
\label{eq:23}
\end{eqnarray}
\medskip

In Eqs.(\ref{eq:21}-\ref{eq:23}), for a given thermodynamical fluctuation $\delta X(\vec{r},t)$, the corresponding Fourier-Laplace transform is denoted as $ \delta \hat{\tilde{X}}=\delta \hat{\tilde{X}}(\vec{q},s)$.

The dispersion relation associated to the system (\ref{eq:21}-\ref{eq:23}) governs the dynamic behavior of the thermodynamic fluctuations, and read:
\begin{equation}
\left|\begin{array}{ccc}
s & n_{o} & 0\\
-k T_{o}q^{2}+4 \pi G m^{2} n_{o}& \tilde{\rho}_{o}s+Aq^{2}+\frac{\kappa T_{o}}{c^{4}} s^{2}& -q^{2}n_{o} k+\frac{\kappa q^{2}}{c^{2}} s\\
0 & k n_{o} T_{o}-\frac{\kappa T_{o}}{c^{2}}s & +c_{n} n_{o} s+\kappa q^{2}\end{array}\right|=0\label{eq:8}\end{equation}
The resulting dispersion relation has the form:
\begin{equation}
a_{0}s^{4}+a_{1}s^{3}+a_{2}s^{2}+a_{3}s+a_{4}=0
\label{eq:disp}
\end{equation}
where
\begin{equation}
a_{0}=\frac{c_{n} n_{o} T_{o} \kappa}{\hat{\rho_{o}} c^{4}}
\label{eq:disp1}
\end{equation}

\begin{equation}
a_{1}=c_{n} n_{o}
\label{eq:disp2}
\end{equation}

\begin{equation}
a_{2}=A c_{n} n_{o} q^{2}-\frac{2 k n_{o} q^{2} T_{o} \kappa}{c^{2}}+ q^{2} \kappa \tilde{\rho_{o}}
\label{eq:disp3}
\end{equation}

\begin{equation}
a_{3}= -4 \pi G M^{2} c_{n} n_{o}^{3}+c_{n} k n_{o}^{2} T_{o} q^{2}+k^{2} n_{o}^{2}T_{o} q^{2}+ A \kappa q^{4}
\label{eq:disp4}
\end{equation}

\begin{equation}
a_{4}=-4 \pi G M^{2} n_{o}^{2} q^{2} \kappa+k n_{o} T_{o}q^{4} \kappa
\label{eq:disp5}
\end{equation}

\medskip
 The gravitational effect only appears in Eqs. (22-23). In this case, the denominator $c^{4}$ in Eq. (19) generates a very large root of the full dispersion relation that can be accurately approximated by solving the simple equation:
\begin{equation}
a_{o} s^{4}+a_{1} s^{3}=0
\label{eq:disp6}
\end{equation}
The non-trivial root of Eq. (\ref{eq:disp6}) is the one obtained in reference \cite{HL1}. It is relevant to notice that, in the absence of dissipation $\kappa=0$, A=0, the dispersion relation reduces to a simple cubic equation that leads to the well-known Jeans wave number.

The conclusion of this analysis is that the heat acceleration coupling would lead to extreme short time scales (of structure formation even in the presence of gravitation $10^{-34}$ seconds for water at 300 K and one atmosphere of pressure \cite{HL1}). This is a serious drawback for Eckart's  coupling Eq. (\ref{eq:17}).
\section{Final remarks}
This work supports the points raised in earlier work \cite{mexmeeting07} where it was shown that the use of the heat-acceleration coupling, introduced since 1940, yields unobserved predictions regarding generic instabilities and light scattering spectra. The results here obtained show that this kind of coupling would also generate a pathological behavior regarding structure formation. Indeed, the introduction of Eq. (\ref{eq:17}) in the balance equations leads to negligible effect of gravity. This is of course in contradiction with  current knowledge. Moreover, the introduction of the so-called second order theories may modify significantly the results here. In most version of those theories the corresponding constitutive equation reads, neglecting viscous effects:
\medskip
\begin{equation}
q^{\nu }+\tau \dot{q}^{\nu }=-\kappa h_{\mu }^{\nu }\left( T^{,\mu }+\frac{T}{c^{2}}\dot{u}^{\mu
}\right)   \label{eq:calabaza}
\end{equation}
\medskip
where $\tau$ is relaxation time much larger the characteristic time of $10^{-34}$ seconds. In this equation the heat acceleration coupling is still retained. On the other hand, a possible first order alternative constitutive equation has been successfully introduced in Ref. \cite{Nos2}. The study of these alternatives regarding the gravitational instability in the presence of these constitutive equations will be addressed in the near future.

The authors acknowledge D. Brun-Battistini for her valuable comments for this manuscript.

\end{document}